 \documentclass[final,5p,times,twocolumn]{elsarticle}

\usepackage{amssymb}
\usepackage{bm}% bold math

\journal{Physica E}

\begin{document}

\begin{frontmatter}



\title{\Large\bf Low-Temperature Resistivity Anomalies in Periodic Curved Surfaces}


\author{Shota Ono and Hiroyuki Shima}

\address{\it Department of Applied Physics, Graduate School of Engineering, Hokkaido University,
Sapporo 060-8628, Japan}

\begin{abstract}
Effects of periodic curvature on the the electrical resistivity of corrugated
semiconductor films are theoretically considered.
The presence of a curvature-induced potential affects the motion of electrons confined to the thin curved film,
resulting in a significant resistivity enhancement at specific values of two geometric parameters:
the amplitude and period of the surface corrugation.
The maximal values of the two parameters in order to observe
the corrugation-induced resistivity enhancement in actual experiments
are quantified by employing existing material constants.
\end{abstract}


\end{frontmatter}


\section{Introduction}
\label{}
Geometric curvature of nano-scale conducting films with curved geometry\cite{prinz1,lorke,prinz2}
provides us a clue to synthesize a new class of quantum devices based on curved nanomaterials.
The most relevant nature is the occurrence of a {\it curvature-dependent potential} (CP)
that acts on low-energy electrons moving in the curved systems\cite{jensen,costa,ike,kaplan,jaffe}.
The mechanism of the CP have been considered from
various perspectives\cite{cant,entin,aoki,fujita,gravesen,marchi,encinosa,ferrari},
which triggered interesting theoretical predictions on the electric transport properties
of curved nanostructures: quantum transmission probabilities in Y-nanojunction\cite{cuoghi},
a charge separation in helicoidal ribbons\cite{atanasov}, and 
an anomalous shift of the Tomonaga-Luttinger exponent in deformed nanocylinders\cite{shima,shima2}
are only a few to mention. 

In the earlier work, the authors
have studied the CP effect on the electrical resistivity of corrugated
semiconductor films\cite{ono}.
A specific magnitude of the corrugation amplitude turned out to cause a significant increase
in the resistivity, which is due to the CP-enhanced population of 
the electron-electron umklapp scattering processes relevant to the resistivity.
This result motivates us to complete a thorough check of the CP effect in a wider range
of two geometric parameters, i.e., the corrugation amplitude and period, that determine
the shape of the corrugated film.
This attempt is crucially important in order to make use of the effect
in developing curved-nanomaterial-based quantum devices.
%%%%%%%%%%%%%%%%%%%%%%%%%%%%    GRAPHICS   %%%%%%%%%%%%%%%%%%%%%%%%%%%%%%%%%%%%%%%%%%%%%%%%%%%%%%%
\begin{figure}
\center
\includegraphics[scale=0.8,clip]{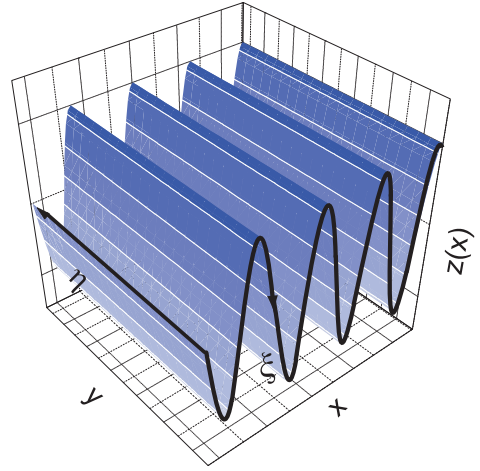}%-----------------------------%
\caption{\label{fig:curve} %(Color online)
Sketch of a periodic curved surface. It is represented by $z=a\cos(\gamma x)$
with a corrugation amplitude $a$ and a period ${2\pi}/{\gamma}$. 
The curvilinear coordinates ($\xi$,$\eta$) used in this paper are embedded in the surface.}
\end{figure}
%%%%%%%%%%%%%%%%%%%%%%%%%%%%%%%%%%%%%%%%%%%%%%%%%%%%%%%%%%%%%%%%%%%%

In this paper, we analyze systematically the correlation between the two geometric
parameters and the CP-induced resistivity enhancement of the corrugated semiconductor films.
We show that the degree of contributions from electron-electron scatterings
to the resistivity is strongly dependent on the values of the two parameters.
The present results enable to determine the maximal surface geometry that is requisite for
the resistivity enhancement to get sizable in actual experiments.
%%%%%%%%%%%%%%%%%%%%%%%%%%%%%%%%%%%%%%%%%%%%%%%%%%%%%%%%%%%%%%%%%%%%%%%%%%%%%%%%%
\section{Gap formation by periodic curvature}
 \label{chap2}
Let us consider a two dimensional (2D) electron system confined to a thin film, in which the film is subjected to
an unidirectional corrugation along the $x$ axis. The height of the film in the $z$ direction is given by
\begin{equation}
 z=a\cos(\gamma x),
\end{equation}
where $a$ and $2\pi/\gamma$ are the amplitude and period of corrugation, respectively (see Fig.~\ref{fig:curve}).
By transforming the variables $x,\ y$ into
\begin{eqnarray}
 \xi =\int_{0}^{x}w(x')dx' ,\ \ \ \eta = y, % \label{eq:xi}
\end{eqnarray}
with the definition $w(x)= \sqrt{1+(\partial z/\partial x)^2 }$,
we obtain the Schr\"{o}dinger equation \cite{ono}
\begin{eqnarray}
 \left[-\frac{\hbar^2}{2m^*}\left(\frac{\partial^2}{\partial \xi^2}+\frac{\partial^2}{\partial \eta^2} \right)
 +U(\xi)\right] \psi (\xi, \eta)=E \psi (\xi, \eta). \label{eq:xieta} 
\end{eqnarray}
Here, the potential $U(\xi)$ is the CP associated with the periodic corrugation,
\begin{equation}
 U(\xi)=-\frac{\hbar^2}{8m^*}
 \frac{[a\gamma^2\cos(\gamma x)]^2}{\left\{1+\left[a\gamma\sin(\gamma x)\right]^2\right\}^{3}}\label{eq:U_xi},
\end{equation}
and $m^*$ is the effective mass of conducting electrons.
Equation (\ref{eq:xieta}) is solved by using the Fourier series expansion
\begin{equation}
 \psi(\xi,\eta)=\left[\sum_{k_\xi} \alpha_{k^\xi} \exp (ik^\xi \xi)\right] \exp (ik^\eta \eta),  \label{eq:wave}
\end{equation}
as a result of which the energy-band structure $E=E(\bm{k})$
with respect to the wavevector $\bm{k}=(k^\xi,k^\eta)$ is obtained.

\begin{figure}
\center
\includegraphics[scale=0.65,clip]{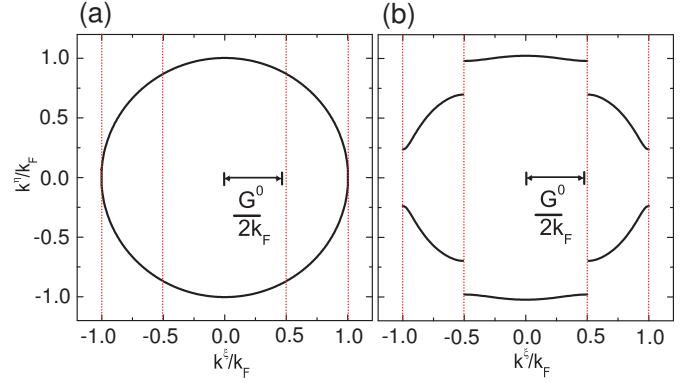}
\caption{\label{fig:fermi}
Fermi circles of the corrugated system with: (a) $\{k_F a, \lambda_F \gamma \}=\{0.7,3.1\}$,
and (b) $\{k_F a, \lambda_F \gamma \}=\{2.7,7.1\}$.
Gaps open at $k^{\xi}/k_F=\pm G^0/(2k_F)$ and $\pm G^0/k_F$ only in Fig.~\ref{fig:fermi}(b).}
\end{figure}

The periodic nature of $U(\xi)$ with the period $\Lambda_\xi=\int_{0}^{\pi/\gamma}w(x)dx$
implies an occurrence of gaps at $k^\xi=\pm nG^0/2$ ($n$ is a positive integer) in the Fermi circle,
where $G^0 \equiv 2\pi/\Lambda_\xi$ defines the reciprocal lattice vector
$\bm{G}\equiv (G^0,0)$ associated with the periodicity of $U(\xi)$.
Such gaps are observed in Fig.~\ref{fig:fermi}(b),
where a sufficiently large amplitude of $U(\xi)$ is assumed by setting
appropriate values of $a$ and $\gamma$.
Still, it should be noted that gaps at $k^\xi=\pm nG^0/2$ may and may {\it not} occur  once $G^0$ is given.
This is primary because $G^0$ as well as $\Lambda_\xi$ are two-variable functions of $a$ and $\gamma$
as understood from Fig.~\ref{fig:curve}.
Hence, different pairs of the two values $\{a, \gamma \}$ can yield the same value of $G^0$, though the
pairs do not always produce a gap at $k^\xi=\pm nG^0/2$.

\begin{figure}
\center
\includegraphics[scale=0.8,clip]{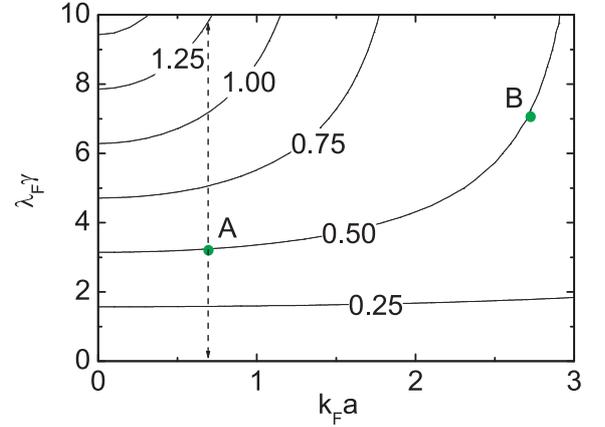}% 
\caption{\label{fig:map}
Contour plot of $G^0/(2k_F)$ as a function of $a$ and $\gamma$.
The points ${\bold A}$ and ${\bold B}$ indicate the numerical conditions to deduce
the gapped (or no-gapped) Fermi circle depicted in Figs.~\ref{fig:fermi}(a) and \ref{fig:fermi}(b), respectively}
\end{figure}

Two contrast examples of the Fermi circle are clearly demonstrated
in Figs.~\ref{fig:fermi}(a) and \ref{fig:fermi}(b).
We set $\{ k_F a, \lambda_F \gamma \}=\{0.7,3.1\}$ and $\{2.7,7.1\}$ to draw the Fermi circles,
in which both pairs give an identical value of $G^0$ as shown in the contour plot of $G^0/(2k_F)$ in Fig.~\ref{fig:map};
the points ${\bold A}$ and ${\bold B}$ in the plot correspond to Fig.~\ref{fig:fermi}(a) and \ref{fig:fermi}(b), respectively.
The Fermi circle in Fig.\ref{fig:fermi}(a) has almost no gap, whereas that in Fig.\ref{fig:fermi}(b) shows
clear gaps at $k^\xi/k_F=\pm G^0/(2k_F)$ and $\pm G^0/k_F$.
As we shall see later, usage of such pairs $\{a, \gamma \}$ that lead to gaps in the Fermi circle is a necessary condition
for the corrugation-induced jump in the resistivity to be observed.
%%%%%%%%%%%%%%%%%%%%%%%%%%%%%%%%%%%%%%%%%%%%%%%%%%%%%%%%%%
\section{Resistivity of the corrugated film}
 \label{umklapp}
We now consider the low-temperature resistivity of 2D electron systems under periodic modulation.
Contributions of two-electron scatterings
to the resistivity at low temperature $T$ are expressed by\cite{ono,uryu}
\begin{eqnarray}%---------RESISTIVITY----------------------------------%
 \rho(T)=\rho_{c}\biggr(\frac{k_BT}{E_F}\biggr)^2 \frac{h}{e^2}, \ \ \ \ \ %\nonumber\\
 \rho_c= \sum_{n=-\infty}^\infty C(n),  \label{eq:rho_c} 
\end{eqnarray}%----------------------------------------------------------%
where $k_B, E_F$ are the Boltzmann constant and the Fermi energy, respectively.
$C(n)$ represent the contributions from the $n$th 
umklapp scatterings $(\bm{k}_1,\bm{k}_2) \rightarrow (\bm{k}_3,\bm{k}_4)$ that satisfy the momentum
conservation law:
\begin{equation}
 \bm{k}_3+\bm{k}_4=\bm{k}_1+\bm{k}_2+n\bm{G}. \label{eq:mom}
\end{equation}
$C(n)=C(-n)$ is satisfied because of the inversion symmetry of the Fermi circle with respect to $k^\xi=0$.
Equation~(\ref{eq:rho_c}) holds for $k_BT\ll E_F$, i.e., when smearing of 
the Fermi degeneracy by thermal fluctuation
can be negligible so that electron-electron scatterings are allowed just on the Fermi circle.
%%%%%%%%%%%%%%%%%%%%%%%%%%%%    GRAPHICS   %%%%%%%%%%%%%%%%%%%%%%%%%%%%%%%%%%%%%%%%%%%%%%%%%%%%%%%
%%%%%%%%%%%%%%%%%%%%%%%%%%%%%%%%%%%%%%%%%%%%%%%%%%%%%%%%%%%%%%%%%%%%%
%%%%%%%%%%%%%%%%%%%%%%%%%%%%%%%%%%%%%%%%%%%%%%%%%%%%%%%%%%%%%%%%%%%%%%%%%%%%
%%%%%%%%%%%%%%%%%%%%%%%%%%%%%%%%%%%%%%%%%%%%%%%%%%%%%%%%%%%%
%%%%%%%%%%%%%%%%%%%%%%%%%%%%%%%%%%%%%%%%%%%%%%%%%%%%%%%%%%%%%%%%%%%%%%%%%%%%
\begin{figure}
\center
\includegraphics[scale=0.8,clip]{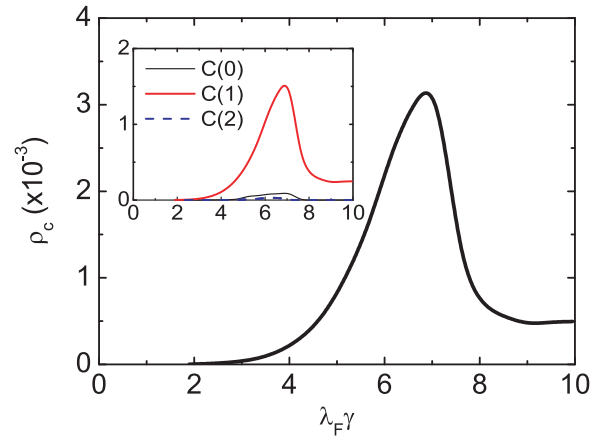}% 
\caption{\label{fig:rho_c1}
Resistivity coefficient $\rho_c$ as a function of $\gamma$ with $k_Fa=0.7$.
$\rho_c$ shows the maximum at $\lambda_F \gamma = 7.0$.
Inset: $n$th-order contributions $C(n)$ for $n=0,1,2$.
The first umklapp process plays a dominant role in determining $\rho_c$.}
\end{figure}
%%%%%%%%%%%%%%%%%%%%%%%%%%%%%%%%%%%%%%%%%%%%%%%%%%%%%%%%%%%%

Each $C(n)$ in Eq.~(\ref{eq:rho_c}) sums up contributions from all possible $n$th-order processes
$(\bm{k}_1,\bm{k}_2) \rightarrow (\bm{k}_3,\bm{k}_4)$
that satisfy Eq.~(\ref{eq:mom}).
Among many processes, specific ones involving the state $\bm{k}$
that locates near the gap are dominant in the total value of $C(n)$\cite{ono}.
Furthermore, the number of such dominant processes
takes the maximum when gaps open at both ends of the Fermi circle ($k^\xi=\pm k_F$),
i.e., when the relation
\begin{equation}
 \frac{G^0}{2k_F}=\frac{1}{n} \ (n=1,2,\cdots), \label{eq:gkf_eq}
\end{equation}
is satisfied. As a result, the surface corrugation with $a$ and $\gamma$ that satisfy  Eq.~(\ref{eq:gkf_eq})
can cause the corrugation-induced jump in $\rho_c$.
We emphasize again that Eq.~(\ref{eq:gkf_eq})
is a necessary (not sufficient) condition of the jump in $\rho_c$, since gaps may
and may not be present for a given $G^0$.
%%%%%%%%%%%%%%%%%%%%%%%%%%%%%%%%%%%%%%%%%%%%%%%%%%%%%%%%%%%%%%%%%%%%%%%%%%%%%
\section{Numerical Results}
 \label{}
In numerical calculations of $\rho_c$, we set $E_F=10$meV and employed material constants
of the GaAs/Al$_{x}$Ga$_{1-x}$As heterostructures:
the effective mass is $m^*/m_0$=0.067 ($m_0$ is a bare electron mass) and
the dielectric constant is $\epsilon /\epsilon_0$=13.2 ($\epsilon_0$ is the dielectric constant of vacuum).

Figure~\ref{fig:rho_c1} shows the $\gamma$ dependence of $\rho_c$;
we fixed $k_Fa$=0.7 and increased $\gamma$ from $\lambda_F \gamma =0$ to $\lambda_F \gamma =10$
as indicated by the dotted trajectory in Fig.~\ref{fig:map}.
$\rho_c$ has a peak at $\lambda_F \gamma =7.0$ under the condition of $k_Fa=0.7$,
which is a consequence of the satisfied relation $G^0/(2k_F)=1$ as seen from Fig.~\ref{fig:map}.
Hence, the peak of $\rho_c$ at $\lambda_F \gamma=$7.0 tells us that the pair $\{a, \gamma \}$
of the above mentioned values generates gaps at $k^\xi=\pm k_F$ in the Fermi circle.
In the inset of Fig.~\ref{fig:rho_c1},
we also show the $\gamma$ dependences of $C(n)$ with $n=0$, 1 and $2$.
The term $C(1)$ is found to be dominant at all values of $\lambda_F \gamma$, which imply that
overall behavior of $\rho_c$ in the $\gamma$ region considered is determined by the first-order umklapp process.

It should be noted that in Fig.~\ref{fig:rho_c1},
no peak appears at $\lambda_F \gamma = 3.1$
although the value satisfies the condition of Eq.~(\ref{eq:gkf_eq}) with $n=2$.
The disappearance of the peak in $\rho_c$ at $\lambda_F \gamma = 3.1$ is attributed
to the absence of gaps associated with the second-order umklapp process, as discussed below in detail.
%%%%%%%%%%%%%%%%%%%%%%%%%%%%%%%%%%%%%%%%%%%%%%%%%%%%%%%%%%%%%%%%
\section{Discussions}
 \label{disc}
\begin{figure}
\center
\includegraphics[scale=0.8,clip]{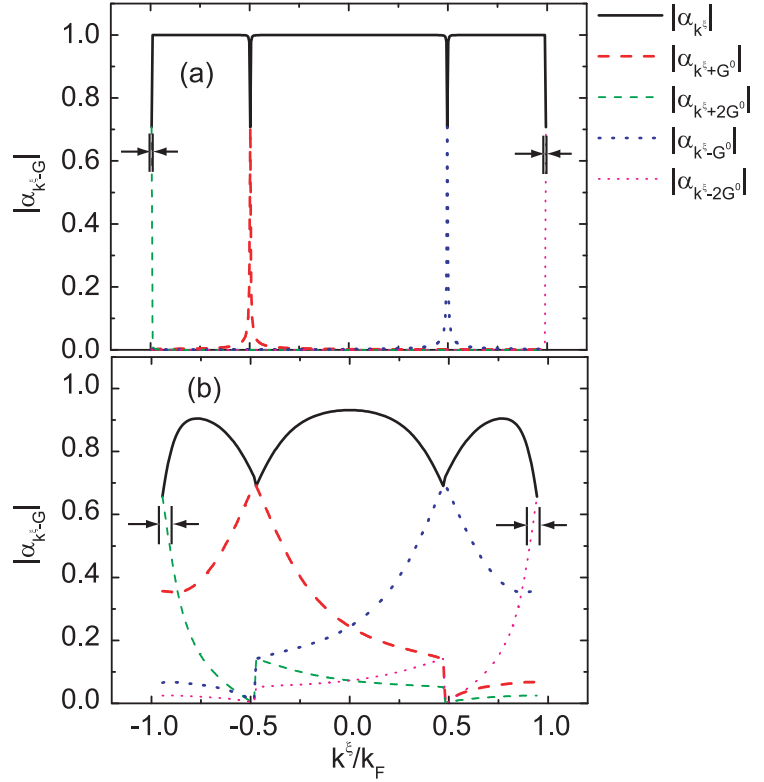}% 
\caption{\label{fig:alpha}
Profile of the Fourier coefficients $\vert \alpha_{k^\xi} \vert$ defined by Eq.~(\ref{eq:wave})
as a function of $k^\xi$. Plots in (a) and (b) correspond to $\{k_F a, \lambda_F \gamma \}=\{0.7,3.1\}$ and
$\{2.7,7.1\}$, respectively. The width of the upward peaks $\Delta k^\xi(n)/k_F$
for $n=2$ are signified by arrows.}
\end{figure}

We discuss the reason why the contributions from the second umklapp processes
do not enhance $\rho_c$ at $\lambda_F \gamma = 3.1$ in Fig.~\ref{fig:rho_c1}.
Figures \ref{fig:alpha}(a) and \ref{fig:alpha}(b) show the profiles of
the Fourier coefficients $\vert \alpha_{k^\xi} \vert$ (see Eq.~(\ref{eq:wave}))
under the numerical conditions of ${\bold A}$ and ${\bold B}$ marked in Fig.~\ref{fig:map}, respectively.
In both figures, upward peaks of $\vert \alpha_{k^\xi \pm nG^0} \vert$ with $n \ne 0$
touch downward peaks of $\vert \alpha_{k^\xi} \vert$ at $k^\xi/k_F=\mp nG^0/(2k_F)$.
The widths $\Delta k^\xi (n)$ of the upward peaks are estimated as \cite{kittel}
\begin{equation}
 \frac{\Delta k^\xi(n)}{k_F}=\frac{k_F}{nG^0}\frac{\vert U_{nG^0} \vert}{E_F}, \label{eq:width}
\end{equation}
where $U_{nG^0}$ is the Fourier component of the CP.
We can deduce from Eq.~(\ref{eq:width}) that $\Delta k^\xi(2)/k_F \sim 10^{-5}$ for ${\bold A}$
and $\sim 10^{-1}$ for ${\bold B}$;
furthermore, it follows from Eq.~(\ref{eq:width}) that
$\Delta k^\xi(2)/k_F$ increases monotonically with increasing $a$ and $\gamma$.
Notice that the larger width $\Delta k^\xi(n)/k_F$ results in the larger $\rho_c$, since the number of
the associated umklapp processes increase.
In fact, $\Delta k^\xi(n)/k_F \ge 10^{-2}$ is needed to get 
$\rho_c \ge 10^{-3}$ under the condition that $\{k_F a, \lambda_F \gamma \}$ yields $G^0/(2k_F)=$1 or 2.
(The magnitude of $\rho_c \sim 10^{-3}$ is large enough to observe it experimentally\cite{messica}.)
These arguments account for the reason why the parameters $\{a, \gamma \}$ corresponding to ${\bold A}$,
which provide $\Delta k^\xi(2)/k_F \sim 10^{-5} \ll 10^{-2}$, give no peak in $\rho_c$ in Fig.~\ref{fig:rho_c1}.

We have also confirmed that $k_Fa\ge 0.5$ for $n=1$ and $k_Fa\ge 2.0$ for $n=2$
are enough to yield $\Delta k^\xi(n)/k_F \ge 10^{-2}$,
where $0<\lambda_F \gamma <10$ is assumed.
Our results reveal the geometric conditions of the surface corrugation to obtain a sizable jump in $\rho_c$
of a semiconductor film. The same approach is applicable to other semiconducting materials
than the GaAs/Al$_{x}$Ga$_{1-x}$As heterostructures we have considered.
%We can, therefore, suggest that the absolute value of the resistivity is controllable easily
%within a range of $\sim 10^{-3}$ in such values $\{a,\gamma \}$.
%%%%%%%%%%%%%%%%%%%%%%%%%%%%%%%%%%%%%%%%%%%%%%%%%%%%%%%%%%%%%%%%%%%%%%%%%%%%%%%%%%%
%%%%%%%%%%%%%%%%%%%%%%%%%%%%%%%%%%%%%%%%%%%%%%%%%%%%%%%%%%%%%%%%%%%%%%%%%%%%
%%%%%%%%%%%%%%%%%%%%%%%%%%%%%%%%%%%%%%%%%%%%%%%%%%%%%%%%%%%%
\section{Conclusion}
\label{sec:level7}
In conclusion, we have demonstrated
the curvature-induced enhancement of $\rho_c$ of periodically curved thin films at low $T$.
Appropriate values of $a$ and $\gamma$ for the resistivity enhancement to appear in measurements
have been clarified using existing material constants.
The umklapp contributions to the resistivity have been found to increase significantly
when the value of $a$ and $\gamma$ satisfy the conditions given by $G^0/(2k_F)=1/n$ and
$\Delta k^\xi(n)/k_F \ge 10^{-2}$. We hope that experimental tests of our theoretical predictions
open a gate for next-generation devices based on curved nanostructures.
%%%%%%%%%%%%%%%%%%%%%%%%%%%%%%%%%%%%%%%%%%%%%%%%%%%%%%%%%%%%%%%%%%%%%%%%%%%%%%%%%%%%%%%%%%%%%%%%%%%%%%%%%%%
\section*{Acknowledgments}
We would like to thank K. Yakubo, S. Nishino, H. Suzuura, and S. Uryu
for useful discussions and suggestions. This study was supported by a Grant-in-Aid for Scientific
Research from the MEXT, Japan. H.S. is thankful for the financial support
from Executive Office of Research Strategy in Hokkaido University.
A part of numerical simulations were carried out using the facilities of the Supercomputer
Center, ISSP, University of Tokyo.
%%%%%%%%%%%%%%%%%%%%%%%%%%%%%%%%%%%%%%%%%%%%%%%%%%%%%%%%%%%%%%%%%%%%%%%%%%%%%%%%%%%%%%%%%%%%%%%%%%%%%%%%

\end{document}